# SUPER-RESOLUTION OF MULTISPECTRAL SATELLITE IMAGES USING CONVOLUTIONAL NEURAL NETWORKS

M. U. Müller, N. Ekhtiari, R. M. Almeida, C. Rieke

UP42 GmbH, Berlin, Germany - (markus.mueller, nikoo.ekhtiari, rodrigo.almeida, christoph.rieke)@up42.com

**ABSTRACT:**

Super-resolution aims at increasing image resolution by algorithmic means and has progressed over the recent years due to advances in the fields of computer vision and deep learning. Convolutional Neural Networks based on a variety of architectures have been applied to the problem, e.g. autoencoders and residual networks. While most research focuses on the processing of photographs consisting only of RGB color channels, little work can be found concentrating on multi-band, analytic satellite imagery. Satellite images often include a panchromatic band, which has higher spatial resolution but lower spectral resolution than the other bands. In the field of remote sensing, there is a long tradition of applying pan-sharpening to satellite images, i.e. bringing the multispectral bands to the higher spatial resolution by merging them with the panchromatic band. To our knowledge there are so far no approaches to super-resolution which take advantage of the panchromatic band. In this paper we propose a method to train state-of-the-art CNNs using pairs of lower-resolution multispectral and high-resolution pan-sharpened image tiles in order to create super-resolved analytic images. The derived quality metrics show that the method improves information content of the processed images. We compare the results created by four CNN architectures, with RedNet30 performing best.

## 1. INTRODUCTION

Super-resolution (SR) is the process of deriving images of higher resolution (HR) by applying an algorithm to a low resolution (LR) image. Single image SR approaches do so by using a single LR image; this is considered a classical problem in computer vision (Dong et al., 2014). Parallel to many other computer vision problems, SR approaches employing deep convolutional neural networks (CNNs) outperformed other techniques over the course of the last few years. SR can be used to improve the results of CNN-based object detection, see e.g. Haris et al. (2018) using RGB photographs and Shermeyer & Etten (2019) for a similar study using satellite imagery.

As deep learning algorithms for super-resolution originated in the computer vision domain, they are primarily developed on RGB images in 8-bit color depth, where the distance from sensor to camera is several meters. When applying these algorithms to satellite images several challenges need to be addressed (Tsagkatakis et al. 2019):

 • Multispectral and hyperspectral data has a higher dimensionality ranging from four to dozens of bands.
 • Analytic satellite image products are calibrated so that they represent a physical unit, either surface reflectance or absolute radiance. These are encoded in 12-bit.[1]
 • Atmospheric conditions, haze, clouds and cloud shadows add additional variation to measured values.
 • Land cover characteristics vary globally to a high degree, a model trained on images of temperate forest areas in Europe might completely fail when applied to images of tropical forests in e.g. South-East Asia.

---
[1] We use the term analytic image in the following to differentiate this type of imagery from visual images (less color depth, converted to RGB 8-bit).

 • Large amounts of data are encoded in each observation because of the large field of view / distance between sensor and scene.

Also, many satellite sensors record image bands in different resolutions, Sentinel-2 images for example have a resolution of 10 m, 20 m or 60 m depending on the band. Very common is the availability of a panchromatic band which achieves high spatial resolution by capturing light in a broader spectral range (limiting information content about reflectance at specific wavelengths). Example datasets are Landsat-8, Satellite Pour l'Observation de la Terre (SPOT) and Pléiades. In the Landsat-8 case the panchromatic band has 15 m, while 8 other bands are at 30 m and two bands at 100 m resolution. The Pléiades and SPOT sensor have four bands at respectively 1.5m/6m, while the panchromatic band has 0.5m/1.5m resolution. The differences in geometric resolution can be leveraged in the super-resolution image enhancement process.

In this work we propose a method for superresolution of Pléiades images by combining CNN-based SR with pansharpening, a classical method of the remote sensing discipline. Our approach is based on state of the art computer vision approaches (Dong et al. 2014, Mao et al. 2016 and Ledig et al. 2017), but extends those by:

1. Processing analytic imagery in its full 12-bit information depth.
2. Super-resolving all four available multispectral bands (Red, Green, Blue and Near-Infrared).
3. Taking advantage of the resolution difference between the multispectral and panchromatic bands for robust training data generation.

The paper is organized as follows: In chapter 2 we discuss related work with regard to pansharpening of satellite imagery and super-resolution of both (RGB-)photographs as well as satellite images. In chapter 3 we introduce our methodology which uses pansharpened 12-bit, 4 band imagery to train a SR CNN. In chapter 4 results of our experiments are described, which include a comparison of the performance of four state of the art SR CNN architectures. Finally, in chapter 5, we evaluate our approach and outline conclusions that can be drawn from our experiments.

## 2. DEEP LEARNING FOR SINGLE IMAGE SUPER-RESOLUTION OF SATELLITE IMAGES

Deriving high resolution from matching lower resolution images has a long history in satellite remote sensing. Many satellite sensors provide multiple data bands at the same, and one additional band at a higher spatial resolution. This panchromatic band is a gray-scale image representing a broader spectral bandwidth which is a result of a trade-off between spatial and spectral resolution. The process of merging multispectral bands with a panchromatic band is called pansharpening, a general definition and review of used methods can be found in Vivone et al. (2015).

The same problem is also tackled in the discipline of Computer Vision; the focus of single-image super-resolution is to derive higher resolution images by algorithmic object enhancements. An early approach to single image super-resolution using CNNs is described in Dong et al. (2014). The approach is motivated by the fact that the sparse-coding-based method (Yang et al., 2008), a traditional SR method, is equivalent to a deep convolutional network. The proposed CNN learns an end-to-end mapping from low- to high-resolution images and is named Super-Resolution Convolutional Neural Network (SRCNN). The model processes three color channels, has a simple architecture consisting of three convolutional layers with different kernel sizes and uses the SGD optimizer (Goodfellow et al., 2016). To synthesize the low-resolution samples, sub-images are blurred by a Gaussian kernel and sub-sampled by the upscaling factor. This approach is used by most CNN-based super-resolution approaches from then on, both for RGB as well as multi- or hyper spectral images. As a preprocessing step of the CNN, these images are upscaled by the same factor via bicubic interpolation. Quality of results is evaluated by using six different metrics including the widely used Peak Signal to Noise Ratio (PSNR) and Structural Similarity (SSIM, Zhou Wang et al., 2004) indices. The model is able to outperform both a bicubic baseline as well as other state of the art models.

SRCNN is applied to satellite imagery by Liebel and Körner (2016) with no changes to the network architecture itself. In comparison to the original paper, the input data is not converted into the YCbCr space, but applied directly to the image bands. The model is re-trained using Sentinel-2 images using the same approach as described in Dong et al. (2014) and afterwards shows better performance than the bicubic baseline. Although only three bands are used in the study, the proof is made that CNN-based super-resolution methods are equally applicable to satellite imagery.

SRCNN was considered slow during execution so a variant called Fast SRCNN (FSRCNN) was developed (Dong et al., 2016). The main change compared to the original architecture is that the separate upsampling step is replaced by a deconvolution layer that is added to the end of the network. A shrinking and expanding layer are also added, resulting in an hourglass-shaped architecture. Finally, Parametric Rectified Linear Unit (PReLU) replaces the SGD optimizer. Performance metrics improve marginally while execution time is reduced by a factor of around 40.

Another variation of SRCNN called Multi-Channel SRCNN (MC-SRCNN) is described in Youm et al. (2016). The key difference between SRCNN and MC-SRCNN is that MC-SRCNN takes multi-channel input images while SRCNN has one single channel input. The multi-channel input is created during a preprocessing step by applying different interpolation algorithms and sharpening filters. The rationale for the approach is that super-resolved images often lack high frequency components, which can be added by the described feature engineering approach. The used metrics PSNR and SSIM improve marginally as a result.

Very Deep Super Resolution (VDSR; Kim et al. 2015) is an alternative approach which uses a very deep convolutional network inspired by VGG-net (Simonyan and Zisserman, 2015). This is the first time a residual network is applied to the problem of image super-resolution. It is found that large depth is necessary for good performance, that the described network converges much faster than SRCNN and also that the same model can be applied to multiple scales. Finally, PSNR and SSIM metrics show improvements across all used input datasets and scales.

The application of autoencoders to a number of image restoration problems including superresolution is discussed in Mao et al. (2016) hereby suggesting another type of CNN architecture. The proposed RED-Net - very deep Residual Encoder-Decoder Network - uses symmetric skip connections between the encoder and decoder sections of the network. Based on this architecture two networks are derived with 20 and 30 layers respectively. The authors conclude that they can achieve better results than other network architectures used so far, among them SRCNN, and that the deeper RedNet30 is the best method overall.

A third major architectural alternative is the Super-resolution Generative Adversarial Network (SRGAN; Ledig et al., 2017). The authors argue that minimizing the mean squared error (MSE) for optimization purposes in super-resolution CNNs will result in good PSNR, but the results lack in high frequency details and are not perceptually pleasing. A residual network called Super Resolution ResNet (SRResNet) is also tested. It is found that SRResNet outperforms all other state of the art approaches, including SRGAN when evaluated using PSNR and SSIM. They also use a Mean Opinion Score (MOS), which evaluates the results based on the judgement of 26 human image quality raters. They conclude that while SRResNet is the start of the art when evaluated with the widely used PSNR measure, SRGAN performs better when the aim is image fidelity.

Applications of super-resolution to satellite images are again addressed in Tuna et al. (2018), Lanaras et al. (2018) and Pouliot et al. (2018). Tuna et al. (2018) apply SRCNN and VDSR to Pléiades as well as SPOT images. Here again only three bands with an 8-bit depth are used, therefore discarding spectral information available from the near infrared band as well as the higher spatial resolution that the panchromatic band would be able to offer. In Lanaras et al. (2018) Sentinel-2 bands of resolution 20 m and 60 m are super-resolved to 10 m using the available high resolution bands to transfer spatial detail to bands of lower resolution. The network architecture called DSen2 is inspired by EDSR (Lim et al., 2017), a ResNet variation. It is to note that in this paper the full 12/16-bit depth of the Sentinel-2 images is taken advantage of. A similar approach is used by Pouliot et al. (2019) with the difference that Sentinel-2 images are used to train a network to improve the spatial resolution of Landsat images. The used model architectures are SRCNN and a deep residual network.

In summary a number of different CNN architectures are applied to the super-resolution problem, starting with a standard CNN (SRCNN), residual networks (VDSR, EDSR, SRResNet), autoencoders (RED-Net) and GANs. When taking the standard metrics PSNR and SSIM into account, the deeper, residual networks deliver the best results.

Kawulok et al. (2019) investigate how performance of deep learning based SR is influenced by how low-resolution training data are obtained. Their study indicates that the training data characteristics have a large impact on the reconstruction accuracy, and the widely-adopted approach of bicubic downsampling of images is not the most effective method for dealing with satellite images. Their study compares results from FSRCNN and SRResNet using different downsampling techniques and conclude that Nearest Neighbour interpolation leads to better performance than bicubic resampling, that the downsampling procedure has tremendous effect on SR performance and that the much deeper architecture of SRResNet does not seem to outperform a relatively simple FSRCNN, when an appropriate downsampling method is used. Overall, they argue that developing better training data preparation routines may be pivotal in making SR suitable for real-world applications.

Shermeyer and Etten (2019) explore the effects of SR techniques for improving object detection algorithm performance in satellite imagery. Using VDSR and a custom Random Forest Super-Resolution framework they find that the application of SR techniques as a pre-processing step provides an improvement in object detection performance at most resolutions with the greatest benefit being achieved at high resolution imagery (30-60 cm) and 4x upsampling.

## 3. EXPERIMENTS

### 3.1 Experiment design

We propose a method for training CNNs aimed at superresolution of multi-band analytic satellite imagery that uses pansharpening to derive HR from LR images. As the panchromatic band of Pléiades data has a 4 times higher resolution than the multispectral bands, the resulting resolution enhancement is also defined by a factor of 4.

In order to derive a training dataset that is representative of different land cover patterns across the world, we chose 40 geographic areas of similar size and derive 5-band analytic Pléiades images which include a panchromatic band. We then apply the smoothing filter-based intensity modulation (SFIM; Liu, 2000) pansharpening method to create the corresponding HR imagery. SFIM has been developed based on a simplified solar radiation and land surface reflection model. By using a ratio between a higher resolution image (panchromatic band) and its low pass filtered (with a smoothing filter) image, spatial details can be modulated to a lower resolution multispectral image without altering its spectral properties and contrast.

Kawulok et al. (2019) found that the quality of CNN-based SR algorithms depends strongly on the method which is used to create LR/HR image pairs and not necessarily on the depth of the used neural network. We therefore compare four neural network architectures spanning from very shallow to very deep networks. We chose the classic SRCNN because it is used as a baseline for all other approaches and is the shallowest described architecture. The second architecture is REDNET in its RedNet30 variation as it represents the autoencoder group of approaches and is among the best performing models. We also implemented an alternative autoencoder architecture which is much shallower than RedNet30 and consists only of convolutional, MaxPool and Residual layers (Figure 1). This architecture was used to find out if we can achieve results comparable to the proven, but very deep architecture of RedNet30 with a simpler and therefore less resource intensive alternative. Finally we selected SRResNet representing a residual network architecture which, based on the metrics PSNR and SSIM, outperforms all other current approaches, but is also very deep. The four chosen architectures therefore represent the most promising CNN architectures that we identified during our literature review (chapter 2), and can also be used to evaluate the value of very deep networks.

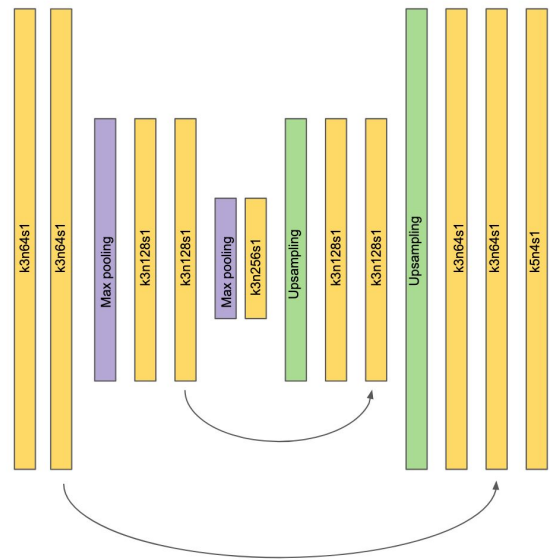

Figure 1. Architecture of the autoencoder network (AESR). Yellow rectangles represent convolutional layers with kernel size k, n feature maps and stride s. Purple rectangles represent max pooling and green rectangles upsampling layers. Arrows indicate skip connections.

As we use 4-band, 12-bit images, it was necessary to extend the original CNN architectures in regard to the dimensions of its tensors. The implementation of the metrics PSNR and SSIM also needed to be updated as the maximum value of an analytic reflectance product is 4095, not 255 as is the case for normal RGB images.

Conventional metrics, such as the peak signal-to-noise ratio (PSNR) operate directly on the intensity of the image, and they do not correlate well with the subjective fidelity ratings. Zhang et al. (2011) found that by incorporating appropriate spatially varying weights, performance quality metrics such as SSIM and PSNR, could be selectively improved and proposed the feature similarity index (FSIM) as an alternative. Aljanabi et al. (2019) build on the concept of FSIM and developed the Information theoretic-based Statistic Similarity Measure (ISSM) which incorporates information theory (Shannon entropy) with a statistic (SSIM), as well as a distinctive structural feature provided by edge detection (Canny). We implemented FSIM and ISSM in Python and made the source code available to the public (https://github.com/up42/image-similarity-measures).

### 3.2 Creation of a training dataset

A set of 40 Pléiades satellite scenes was selected across the globe, making sure land cover types are uniformly distributed e.g. tropical forests, urban environments, desert areas (Figure 2). From each of these scenes, a smaller subset of a representative area of a particular scene with 2 x 2 km dimension was selected.

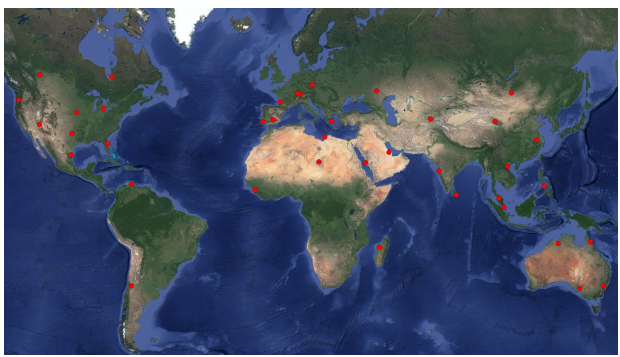

Figure 2. Geographical distribution of 40 areas selected for gathering training data.

The multispectral imagery (low resolution, 2 m) was then pansharpened (high resolution, 0.5 m) and subsequently tiled into 128x128 pixel and 32x32 pixel tiles for the pan-sharpened and multispectral image respectively. In addition, an augmentation procedure was applied along with the tiling, making use of starting points for the tile in ½ of the previous tile in both the horizontal and vertical dimension. This resulted in a data set of 91,968 tiles, on which a 80/20 training/validation random split was performed.

### 3.3 Quantitative results

|      | SR     | AE     | RedNet     | ResNet | Bicubic |
|------|--------|--------|------------|--------|---------|
| PSNR | 31.817 | 31.864 | 31.884     | 31.659 | 31.926  |
| SSIM | 0.6650 | 0.6651 | **0.6653** | 0.6526 | 0.6228  |
| FSIM | 0.4967 | 0.4976 | **0.4977** | 0.4914 | 0.4969  |
| ISSM | 0.0824 | 0.0913 | **0.0920** | 0.0845 | 0.0889  |

Table 1: Comparison of PSNR, SSIM, FSIM and ISSM values for the four implemented models and bicubic interpolation. In bold the value indicating the best performing model.

We selected eight images covering different land cover classes (independent and not contained in the training data) for validation of the results (first column in Appendix A). Super-resolved images derived from multispectral data using the four models were compared with their pan-sharpened image equivalents. PSNR, SSIM, FSIM and ISSM values for the four models, as well as simple bicubic interpretation that can be used as a baseline can be found in Table 1.

### 3.4. Visual interpretation of results

While indices such as PSNR and SSIM are essential performance metrics for super-resolution, the visual comparison between the original image and results is equally important. CNN-based object detection algorithms mostly rely on clearly defined contour shapes to perform well. In Appendix A results from the four CNN approaches when applied to the eight validation images can be seen. In these examples the super-resolution algorithms were applied to the multispectral images and then used to compute the metrics in Table 1. Image quality of the super-resolved images cannot match the pan-sharpened versions, but when compared to the multispectral versions, improvements in spatial detail are clearly visible. At the displayed scale, the alternatives SRCNN, AESR and Rednet are nearly undistinguishable, but the ResNet results display artifacts and a slight loss in colour depth.

As the use cases for super-resolution of satellite images are mostly connected to object detection we additionally processed a set of three images showing a plane, buses and ships, respectively. In Appendix B the three images are displayed in their original multispectral resolution, after pansharpening and application of super-resolution using the best performing model, Rednet30. We want to stress that in Appendix A super-resolution was applied to the multispectral images, while in Appendix B the images were first pansharpened and then super-resolved. In all three examples it is only after pansharpening is applied that the actual objects can be identified unambiguously. The super-resolution step adds an additional level of object definition. The outlines of the objects, be it planes, buses or ships are much more well-defined. In the case of the buses (second row) additional detail on the roofs can be identified, while in the case of the ships this effect can be discerned especially well on the white ship in the upper right of the image.

### 4. DISCUSSION

In this work we introduce a new method for super-resolution of multispectral satellite images that takes advantage of the panchromatic band for creating training data. The method is based on state-of-the-art convolutional neural networks, but is able to process images including a near-infrared band and uses the full 12-bit depth of the data.

We implement four different neural network architectures and compare the results using the well established PSNR and SSIM metrics, the more sophisticated metrics FSIM and ISSM as well as visually. The results clearly indicate that the approach is successful with the overall best performing model being RedNet30, an autoencoder which also uses residual blocks, measured by all metrics besides PSNR. A shallower, but similar architecture that we call AESR performs almost as well. The shallow and simple SRCNN still produces better results than the very deep residual network ResNetSR. In many other studies ResNetSR is the best performing model overall, which indicates that its success depends on the usage of a standard approach to create training data using downsampled images. This conclusion is in line with the findings of Kawulok et al. (2019).

It is interesting to note that all four models perform slightly worse than simple bicubic resampling when measured using PSNR, but better according to SSIM values. Additionally it can be seen that SRCNN and ResNetSR produce higher SSIM but lower FSIM and ISSM values than bicubic interpolation. As FSIM and ISSM were developed as an improvement to error-based approaches such as PSNR and SSIM it can be concluded that our autoencoder based networks AESR and RedNet30 lead to a genuine image improvement and are superior to simple interpolation.

One possibility to further improve model performance would be to use one of the more complex and better performing approaches for pan-sharpening during the training data creation process. According to Vivone et al. (2015) a number of algorithms produce better results than the SFIM method that we used, with the overall best algorithm being band-dependant spatial detail (BDSD).

As is true for all deep learning based models, using additional and even more diverse training data might also lead to an improvement of model performance.

## 5. CONCLUSION

In this research we developed and evaluated a new approach for super-resolving multispectral, analytic satellite images combining CNNs with training data generation based on pansharpening. We found that deep neural networks combining the autoencoder approach combined with skip connections produces the best results.

To our knowledge this is the first time a CNN-based super-resolution method is implemented that both (1) takes advantage of the higher resolution provided by the panchromatic band and (2) is able to process 12-bit multispectral data.

The method can be improved further by creating larger and higher quality training data e.g. by using alternative methods for pansharpening. It could also be adapted to process datasets consisting of bands with different resolutions e.g. Sentinel-2 images with their 10 m, 20 m and 60 m resolution bands.

**APPENDIX**

Appendix A. Eight validation images pan-sharpened, processed by all CNN alternatives and using bicubic upsampling (imagery ©Airbus Defence and Space 2020). Pansharpening and super-resolution was applied to the multispectral images. Rows 1-8: Airplanes at an airport in Berlin, Germany (1), forest border in Kalimantan, Indonesia (2), harbour storage structures in Singapore (3), country road near Tehran, Iran (4), a farm near Lafayette, USA (5), buildings in Tokyo (6), rice paddies near Ubud, Bali (7) and a parking lot in Auckland, New Zealand (8).

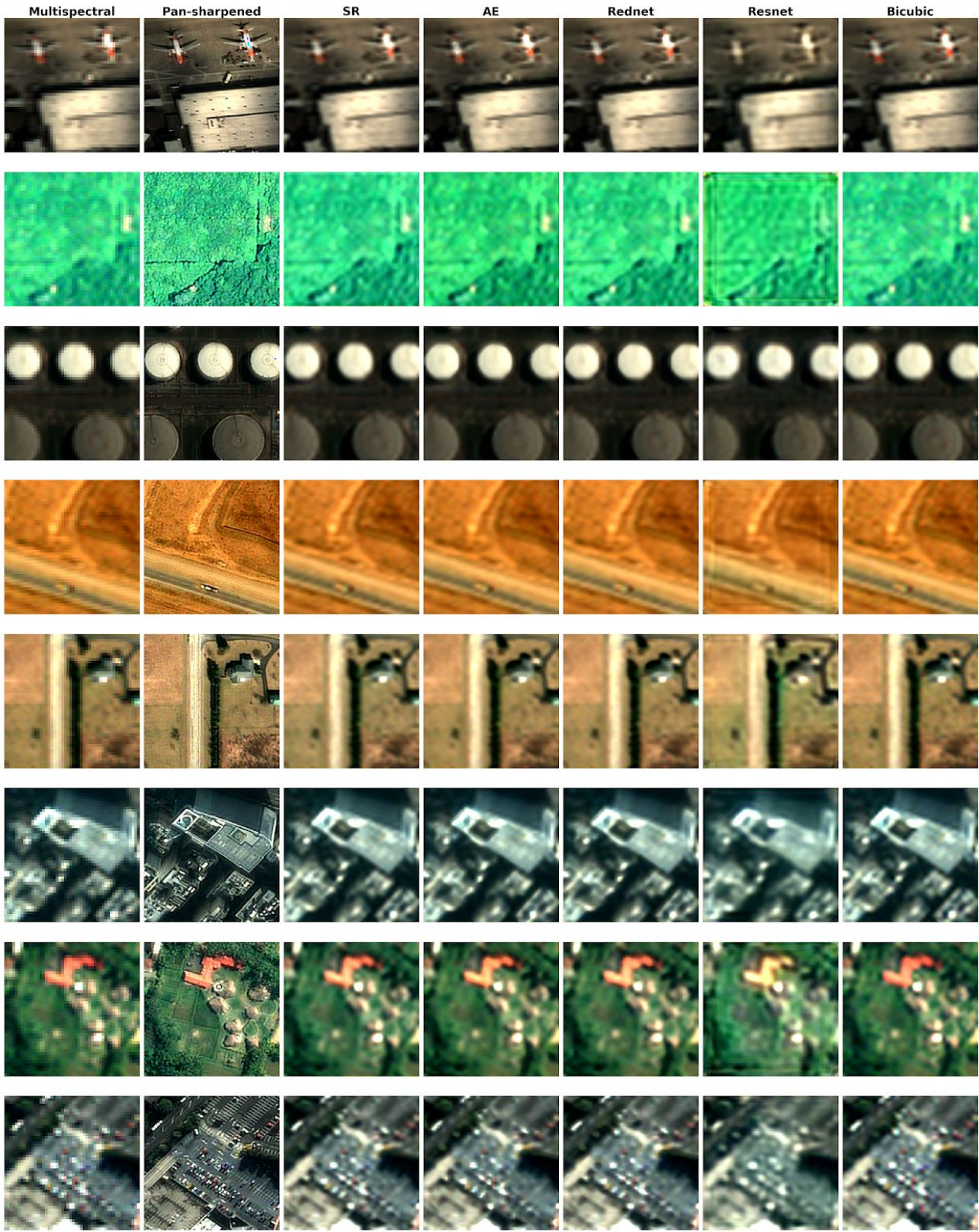

Appendix B. Image examples for the original image, pansharpened and super-resolved results (from left to right) (imagery ©Airbus Defence and Space 2020). Super-resolution was applied to the pan-sharpened images to show how the final, resulting images would appear. Rows 1-3: Airplanes at an airport in Tucson, USA; buses in Tehran and boats in the harbour of Capetown.

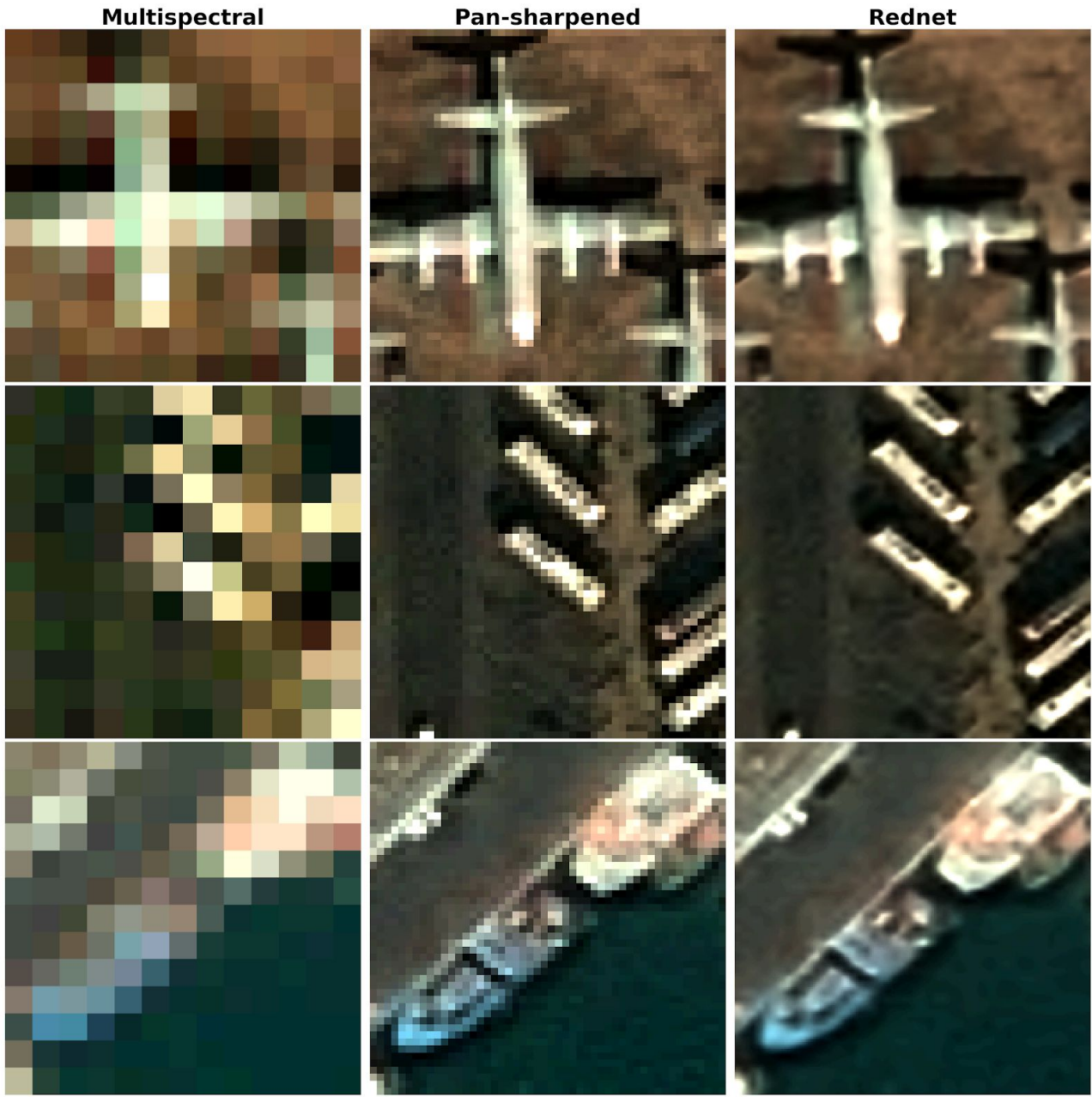